\documentclass[aps,prl,twocolumn,showpacs,superscriptaddress,groupedaddress]{revtex4-2}  % for review and submission
\usepackage{graphicx}  % needed for figures
\usepackage{dcolumn}   % needed for some tables
\usepackage{bm}        % for math
\usepackage{amssymb}   % for math
\usepackage{amsmath}   % for math

\usepackage{verbatim}  % comments 
\usepackage{xcolor}    % color of text

\newcommand{\egi}[1]{{\color{red} #1}}% corrections

\begin{document}
\title{Thermal-energy relaxation in chiral quantum Hall edge mode}

\author{Edvin G. Idrisov}
\affiliation{Department of Physics, United Arab Emirates University, P.O. Box 15551 Al-Ain, United Arab Emirates}

\date{\today}

% The following information is for internal review, please remove them for submission
%\widetext
%\leftline{Version xx as of \today}
%\leftline{Primary authors: Joe E. Physics}
%\leftline{To be submitted to (PRL, PRD-RC, PRD, PLB; choose one.)}
%\leftline{Comment to {\tt d0-run2eb-nnn@fnal.gov} by xxx, yyy}
%\centerline{\em D\O\ INTERNAL DOCUMENT -- NOT FOR PUBLIC DISTRIBUTION}

% the following line is for submission, including submission to the arXiv!!
%\hspace{5.2in} \mbox{Fermilab-Pub-04/xxx-E}

%\title{Template for PRL/PRD Papers}
%\input author_list.tex       % D0 authors (remove the first 3 lines
                             % of this file prior to submission, they
                             % contain a time stamp for the authorlist)
                             % (includes institutions and visitors)
%\date{\today}

\begin{abstract}
We study the nonequilibrium state generated by a real Ohmic contact (ROC) in the charge-fractionalization regime. We then investigate its thermal-energy relaxation along a quantum Hall edge mediated by a fictitious Ohmic contact (FOC) with large capacitance and perfect coupling to the edge. For cold Ohmic contacts, the electron distribution function and current noise are fully described by their equilibrium forms and the corresponding Fermi–Dirac distribution. In contrast, in the case of hot Ohmic contacts, strong thermalization effects do not drive the outgoing state into complete equilibrium. Both the distribution function and the noise retain signatures of the initial nonequilibrium state and therefore cannot be characterized solely by the effective contact temperature, contrary to common assumptions.
\end{abstract}

\maketitle

%\section{\label{sec:level1}First-level heading}
% sections are not used for PRL papers

The experimental progress in hybrid mesoscale systems based on chiral QH edge states has triggered more accurate theoretical studies of phase coherence~\cite{Roche1,Bisognin2019,Pierre1,Heiblum0,Nakamura2020}, charge and heat quantization~\cite{Jezouin2016,IdrisovThermalDecay,ArturEquil,Sivre2018,Fujisawa1,Sivre2019}, relaxation and
equilibration~\cite{Altimiras2010,Chalker,IvanRelaxation,Slobodeniuk1,Fujisawa2,Fujisawa3,Rodriguez2020}, and entanglement~\cite{Grenier,Jullien,Bauerle,Fletcher2019}, which are fundamental questions of mesoscopic physics. Among them, the investigation of the electron distribution function and equilibration along the QH edge states, for example, the unresolved issue of heat leakage~\cite{AltimirasRelaxation,Sueur}, remains an open question. In these experiments a quantum point contact (QPC) serves as an essential element. Namely, one prepares a non-equilibrium double-step distribution function with the help of biased partition QPC. Later, the evolution of distribution function along edge states is studied by measuring the conductance of tunneling current between two edges using the quantum dot or the second QPC as a detector~\cite{Ivanbook}. 

Further progress in experimental techniques allowed experimentalists to realize systems, where the QH edge states are perfectly coupled to Ohmic contact, a mesoscale piece of metal with zero level spacing and finite charging energy, $E_{\rm c}=e^2/2C$, where $C$ is the capacitance~\cite{Jezouin}. It is worth mentioning that the Ohmic contact has become a key element of new experimental and theoretical research~\cite{Zubairbook}. The remarkable examples are the suppression of charge quantization caused by quantum fluctuations both in quantum and thermal regime~\cite{Jezouin2016,IdrisovThermalDecay}, heat Coulomb blockade effect~\cite{ArturEquil,Sivre2018,Sivre2019}, interaction induced recovery of the phase coherence~\cite{Clerk,IdrisovDephasing,Pierre0}, charge Kondo effect~\cite{Iftikhar2015,Iftikhar2018}, quantization of the anyonic heat flow~\cite{Heiblum1}, observation of the half-integer thermal Hall conductance~\cite{Heiblum2,Mross}, the measurement of full counting statistics of electron currents with the help of quantum ammeter~\cite{QuantumAmmeter}, and tunneling effects in multisite quantum Hall circuits~\cite{GoldHaber,KarkiOne,KarkiTwo}. Recently, an implementation of Ohmic contact with chiral edge channels allowed experimentalists to investigate relaxation and equilibration, measuring zero frequency noise for both integer and fractional filling factors~\cite{Rosenblatt}. This is the alternative method compared to conventional conductance measurements with respect to applied bias~\cite{Ivanbook}. Subsequently, it has been used to study quantized heat flow in the Hofstadter butterfly~\cite{Hofstadter}, the ballistic-to-diffusive transition in quantum Hall channels~\cite{BalDiff}, dynamical Coulomb blockade under a temperature bias~\cite{DynCoulomb}, and topological effects in thermal Hall conductance~\cite{TopoHeiblum}. In addition to quantum Hall systems based on two-dimensional electron gas, this experimental technique has been used for other two-dimensional materials, including graphene, to investigate heat equilibration between integer and fractional quantum Hall edge modes~\cite{GrapheneEquil} and heat transport in Moiré Chern bands of bilayer graphene~\cite{DasOne,DasTwo}.

\begin{figure}
\includegraphics[width=\columnwidth]{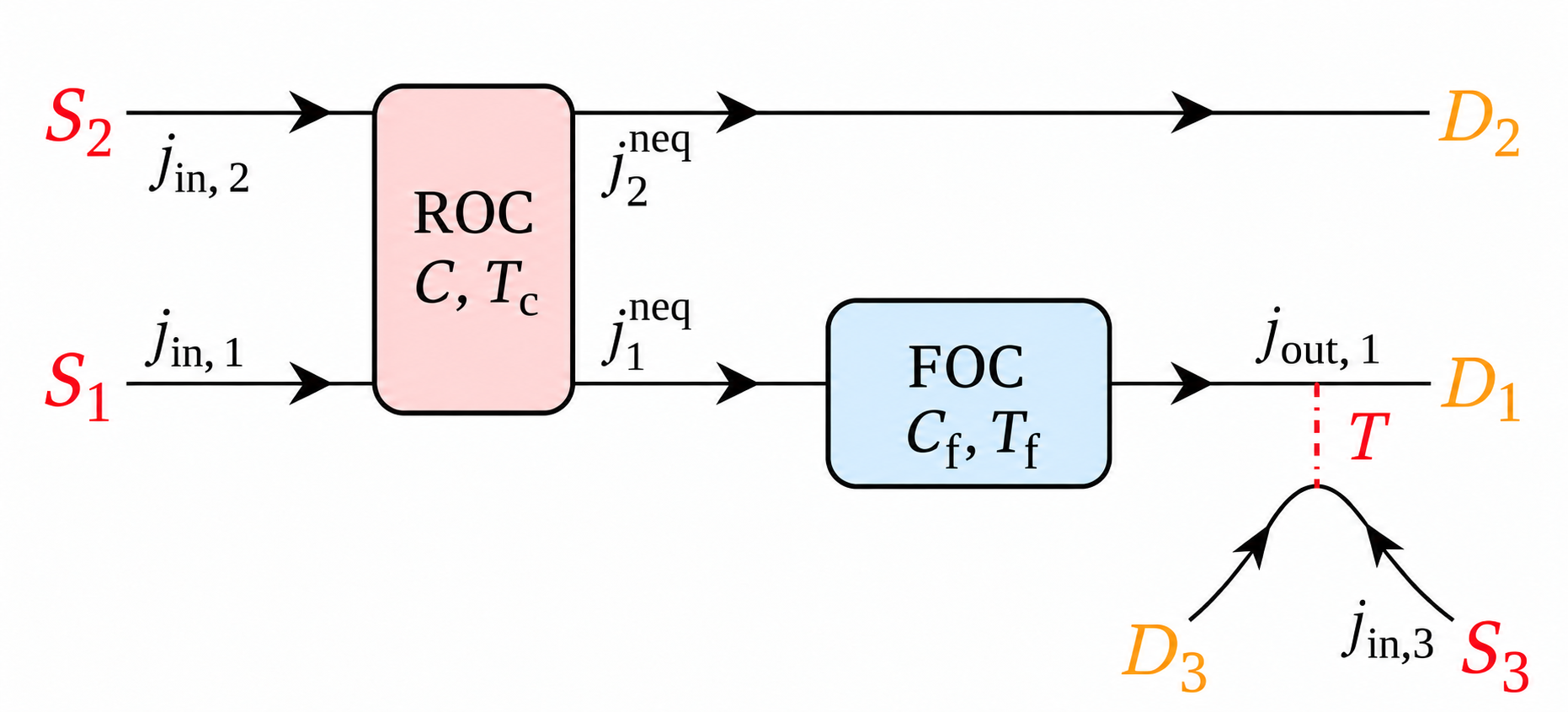}
\caption{\label{fig:one} Schematic of setup. It  consists of the real Ohmic contact (left), a micrometer size piece of metal with negligible level spacing and capacitance $C$. The real Ohmic contact is used to prepare the initial non-equilibrium distribution function with the help of two biased sources $S_1$ and $S_2$. Relaxation of initially prepared state is modeled by a fictitious Ohmic contact with large capacitance $C_{\rm f}$, located after real Ohmic contact.  The setup can be used to measure the  distribution function, $f_{\rm out,1}(\epsilon)$, and noise, $S_{\rm out,1}(\omega=0)$, right before and after passing the region modeled by fictitious Ohmic contact.}
\end{figure}

In this Letter, we study the energy relaxation by Ohmic contact strongly coupled to the QH edge state at filling factor $\nu=1$ (see Fig.~\ref{fig:one}). The setup consists of a real Ohmic contact (ROC), which at zero capacitance ($C \to 0$) serves as an ideal electron fractionalizer. Two independently biased current sources, $S_1$ and $S_2$ with base temperature $T_{\rm in}$ inject current to the ROC. The sources biases, $V_1=-V_2=V$, are chosen such that the total accumulated charge in the ROC vanishes, namely $\langle Q_{\rm c}\rangle=(V_1+V_2)C/2R_q=0$, while the charge fluctuations remain nonzero, where $R_q=2\pi \hbar/e^2$ is the quantum resistance. The equilibration in the ROC increases it's temperature $T_{\rm c} > T_{\rm in}$ with non-zero dissipated power. Further, the measured distribution function and noise allow the study of the relaxation that occurs after the ROC. Relaxation is modeled by a fictitious Ohmic contact (FOC) with capacitance $C_{\rm f}$ and negligible level spacing. The main interest is the study of relaxation modeled by the FOC with a large capacitance. Throughout the paper, we set $|e|=\hbar=k_B=1$. 
%==========================================================================

\textit{Model and formulation of the problem.} We use effective field theory~\cite{Wen} and the bosonization technique~\cite{Giamarchi} to describe the QH edge states perfectly coupled to the Ohmic contact. According to effective theory, the collective fluctuations of the charge densities
$\rho_{\alpha}(x,t)$ and currents $j_{\alpha}(x,t)$ in the QH edge channels are expressed in terms of the bosonic fields $\phi_{\alpha}(x,t)$,
\begin{equation}
\label{Bosonization. Definition of charge density and current}	
\rho_{\alpha}=(1/2\pi)\partial_x \phi_{\alpha},\quad
j_{\alpha}=-(1/2\pi)\partial_t \phi_{\alpha},
\end{equation} 
where the index $\alpha$ denotes the incoming and outgoing channels (see Fig.\ \ref{fig:one}). As stated in the bosonization technique, the electrons in these channels are represented by the vertex operators $\psi_\alpha(x,t)\propto e^{i\phi_\alpha(x,t)}$, so that the electron distribution function in the outgoing channel right after real or fictitious Ohmic contact reads
\begin{equation}
\label{Definition of distribution function}
\begin{aligned}	
& f_{\alpha}(\varepsilon) = \int dt e^{-i\varepsilon t} K_{\alpha}(t), \\
& K_{\alpha}(t) \propto \langle e^{-i\phi_{{\alpha}}(t)}e^{i\phi_{\alpha}(0)}\rangle.
\end{aligned}
\end{equation}
The prefactor in the expression for $K_{\alpha}(t)$ can be fixed in the end of the calculations by comparing with the equilibrium correlation function for free electrons. In particular, the equilibrium two-point correlation function of the incoming states $K^{\rm eq}_{\rm in}(t)=-iT_{\rm in}/ \sinh[\pi T_{\rm in}(t-i0)]$ results in a Fermi-Dirac distribution function $f_{\rm FD}(\epsilon/T_{\rm in})=1/(1+e^{\epsilon/T_{\rm in}})$. Apart from defining the electron distribution function, one can study the non-symmetrized noise of outgoing current after ROC and FOC,
\begin{equation}
\label{Definition of finite frequency noise}
S_{\rm out}(\omega)= \int dt e^{i\omega t} \langle \delta j_{\rm out}(t) \delta j_{\rm out}(0)\rangle,
\end{equation}
where $\delta j_{\rm out}(t)=j_{\rm out}(t)-\langle j_{\rm out}(t)\rangle$ and the average is taken with respect to the incoming states. The symmetrized noise is defined as the sum of the positive and negative frequency components of the non-symmetrized noise, $[S_{\rm out}(\omega)+S_{\rm out}(-\omega)]/2$, which
accounts for fluctuations associated with both emission and absorption processes and is an even function of frequency. At zero frequency, the non-symmetrized and symmetrized noises coincide. It is important to note that, in general, one cannot apply the relation 
\begin{equation}
\label{Wrong noise}
S_{\rm out}(0)=G_{\rm q}\int d\epsilon [1-f_{\rm FD}(\epsilon/T_i)]f_{\rm FD}(\epsilon/T_i),	\quad i={\rm c,f},
\end{equation}
for zero frequency noise of current right after the ROC and FOC, since Wick's theorem is invalid for the fermions outgoing from both Ohmic contacts. It will be shown later that this behavior is related to the nonequilibrium character of the outgoing state prepared by the ROC; due to the presence of electron-electron interactions between two incoming states and strong thermalization effects inside the ROC. In Eq.~(\ref{Wrong noise}), $G_{\rm q}=1/R_{\rm q}=e^2/2\pi \hbar$ is the quantum conductance, and $T_{\rm c}$ is the temperature of the ROC and $T_{\rm f}$ is the temperature of the FOC, which are obtained in Eqs.~(\ref{Real Ohmic contacts temperature}) and (\ref{Fictitious Ohmic contacts temperature. Large capacitance}) below. These temperatures are different from $T_{\rm in}$ due to thermalization effects.

%================================================================================
\textit{Langevin equations and noise in the outgoing channel.} We apply the Langevin equation method to evaluate the electron correlation function $K_{\alpha}(t)$ in Eq.~(\ref{Definition of distribution function}). According to this method, the Ohmic contact serves as a reservoir of neutral excitations, which are accounted for by introducing Langevin currents $j_{{\rm c},l}$ into the equation of motion for the accumulated charge at the contact. The Langevin equations can be obtained rigorously using the bosonization technique for dynamics of bosonic fields~\cite{ArturEquil,IdrisovDephasing,IdrisovThermalDecay,QuantumAmmeter,IdrisovThermalDrag,IdrisovLast}. In the frequency domain, the outgoing currents immediately after the ROC are expressed in terms of the incoming currents and Langevin sources
\begin{equation}
\label{Solution to outgoing currents}
j^{\rm neq}_1(\omega)=j_{{\rm c},i}(\omega)+\sum_{l=1,2} \mathcal{T}(\omega)\left[j_{{\rm in},l}(\omega)-j_{{\rm c},l}(\omega)\right],
\end{equation}
where the subscript $i=1,2$ labels the channel and $\mathcal{T}(\omega)=1/(2-i\omega R_{\rm q} C)$ is the scattering coefficient. The superscript on the left hand side denotes that the outgoing state is non-equilibrium due to thermalization effects and charging energy. The first term in the denominator of the scattering coefficient, $\mathcal{T}(\omega)$, counts the number of incoming and outgoing channels with respect to ROC. Similar equations can be written in the presence of additional FOC. The outgoing current after  FOC has the form  
\begin{equation}
\label{Outgoing current after Nth fictitious contact}
j_{\rm out,1}(\omega)=\mathcal{A}(\omega) j^{\rm neq}_1(\omega)+\mathcal{B}(\omega) j_{{\rm f},l}(\omega),
\end{equation}
where the scattering coefficients are given by $\mathcal{A}(\omega)=1/(1-i\omega R_{\rm q} C_{\rm f})$ and $\mathcal{B}(\omega)=1-\mathcal{A}(\omega)$ and $C_{\rm f}$ is the capacitance of FOC. For the case of no FOC, $C_{\rm f} \to 0$, and one has $j_{\rm out,1}=j^{\rm neq}_1$. The two-point correlation functions of the incoming currents and Langevin sources according to the fluctuation-dissipation theorem are given by the equilibrium spectral function~\cite{Landau}
\begin{equation}
\label{DFT for currents}
\langle \delta j_{k,l}(\omega) \delta j_{k^{\prime},l^{\prime}}(\omega^{\prime})\rangle=2\pi \delta_{k k^{\prime}} \delta_{l l^{\prime}} \delta(\omega+\omega^{\prime}) S_{k,l}(\omega),
\end{equation}
where $S_{k, l}(\omega)=G_{\rm q} \omega/(1-e^{-\omega/T_{k, l}})$ and $k={\rm in, c, f}$ and $l=1,2,3$. These correlation functions are used to determine the temperatures of the ROC and FOC, as well as to calculate the distribution function and noise.

%===============================================================================
\textit{Heating effects in Ohmic contacts.} Before calculating the distribution function and noise, one needs to take into account the thermalization effects inside the Ohmic contacts, i.e., one needs to find the temperatures $T_{\rm c}$ and $T_{\rm f}$ in Eq.~(\ref{DFT for currents}). To do this, we solve self-consistently the energy balance equations, i.e. conservation of incoming and outgoing heat currents, where the heat current is defined by Joule's law, $J_{\alpha}=R_{\rm q} \langle j^2_{\alpha}\rangle/2$.

Further, using Joule's law, we obtain the equation determining the ROC temperature, $V^2/R_{\rm q}+(\pi T_{\rm in}^2/12)\left[2+\mathcal{F}(\pi T_{\rm in}R_{\rm q}C/2)\right]=(\pi T_{\rm c}^2/12)\left[2+\mathcal{F}(\pi T_{\rm c}R_{\rm q}C/2)\right]$, where $\mathcal{F}(x)=(3/x^2)\left[x+\ln(2x)+\Psi(1/2x)\right]$ is a dimensionless function, and $\Psi(x)=d\ln\Gamma(x)/dx$ denotes the digamma function and $\Gamma(x)=\int_0^\infty y^{x-1}e^{-y}\,dy$ is the gamma function. In the absence of FOC and in the small-capacitance regime, $T_{\rm c}R_{\rm q}C\ll 1$, the dimensionless function approaches $\mathcal{F}(x)\to -1$. Therefore, the ROC temperature is given by ~\cite{supp} 
\begin{equation}
\label{Real Ohmic contacts temperature}
T^2_{\rm c} = T^2_{\rm in}+6V^2/\pi^2.
\end{equation}
This equation expresses the heating of the ROC as a result of thermalization. Incoming channels initially have the base temperature $T_{\rm in}$, while the symmetric voltage bias, $V_1=V_2=-V$, introduces an additional nonequilibrium energy proportional to $V^2$. Furthermore, electron–electron interactions within the ROC redistribute this excess energy among the neutral modes $j_{{\rm c}, l}$, converting it into thermal energy and producing the temperature $T_{\rm c}$, which is indeed higher than the base temperature.   

For the FOC, a similar analysis in the large-capacitance regime, \(T_{\rm in}R_{\rm q}C_{\rm f}\gg 1\), yields the following expression for the fictitious contact temperature~\cite{supp} 
\begin{equation}
\label{Fictitious Ohmic contacts temperature. Large capacitance}
T^2_{\rm f} = T^2_{\rm in}+3V^2/\pi^2.
\end{equation}
Compared with the ROC, the FOC reaches a lower temperature because it receives only part of the excess energy generated by the two biased incoming channels; because it is coupled to one edge channel. The heating effects discussed here are used below for the calculation of the distribution function and noise. At $V=0$ the system is in equilibrium and at $T_{\rm f}=T_{\rm c}=T_{\rm in}$.

\textit{Noise in the outgoing channel.} To find the outgoing noise after the FOC, we use Eq.~(\ref{Outgoing current after Nth fictitious contact}) and arrive at the following equation
\begin{equation}
\label{Outgoing noise. Main text result}
S_{\rm out, 1}(\omega)=S^{\rm eq}_{\rm f}(\omega)+|\mathcal{A}(\omega)|^{2}\left[S^{\rm neq}_{\rm in,1}(\omega)-S^{\rm eq}_{\rm f}(\omega)\right],
\end{equation}
where $S^{\rm eq}_{\rm f}(\omega)$ is an equilibrium spectral function at temperature $T_{\rm f}$ from Eq.~(\ref{DFT for currents}). The non-equilibrium incoming noise, $S^{\rm neq}_{\rm in,1}(\omega)$, in the case of fractionalizer (after ROC), when $C \to 0$ is given by~\cite{supp} 
\begin{equation}
\label{Non-equilibrium finite frequency noise for real Ohmic contact in fractionalizer regime. Main text}
S^{\rm neq}_{\rm in,1}(\omega)=\left[S^{\rm eq}_{\rm in}(\omega)+S^{\rm eq}_{\rm c}(\omega)\right]/2.
\end{equation} 
Furthermore, at zero frequency and $\omega R_{\rm q} C_{\rm f} \to 0$ and $\omega /T_{{\rm in,c}} \to 0$ we have $|\mathcal{A}(\omega)|^{2} \to 1$; thus, the outgoing noise is given by the incoming non-equilibrium noise~\cite{supp}, namely,
\begin{equation}
\label{Zero frequency noise for outgoing state with N Ohmic contacts}	
S_{\rm out, 1}(0)=S^{\rm neq}_{\rm in,1}(0)=
G_{\rm q}(T_{\rm in}+T_{\rm c})/2.
\end{equation}
From the above equation, one can see that in the case of cold ROC, i.e. $T_{\rm c}/T_{\rm in} \to 1$, the noise is given by the equilibrium result $S_{\rm out, 1}(0) = G_{\rm q} T_{\rm in}$. However, in the hot ROC limit, $T_{\rm c}/T_{\rm in}\to\infty$, the result deviates from the equilibrium expectation at temperature $T_{\rm c}$. Instead, the zero-frequency outgoing noise is given by $S_{\rm out,1}(0)=G_{\rm q}T_{\rm c}/2$. Thus, Eq.~(\ref{Wrong noise}) does not provide the correct result in this regime, since it assumes an equilibrium Fermi–Dirac distribution with temperature $T_{\rm c}$, yielding $S_{\rm out,1}(0)=G_{\rm q}T_{\rm c}$. It means that the relation for noise obtained using Wick's theorem is not applicable, since the outgoing state is highly non-equilibrium and keeps the information about ROC, where the strong electron-electron interaction between two incoming states is present. However, in the limits $\omega R_{\rm q}C_{\rm f}\to\infty$ followed by $\omega/T_{\rm in,c}\to 0$, one obtains $S_{\rm out,1}(0)=G_{\rm q}T_{\rm c}$. Nevertheless, the zero-frequency noise of the nonequilibrium "in,1" outgoing ballistic channel  downstream of the ROC remains described by Eq.~(\ref{Zero frequency noise for outgoing state with N Ohmic contacts}).

\textit{Non-equilibrium distribution function.} As a first step, we consider the prepared non-equilibrium state after the injection source, i.e. ROC. Our interest is the regime of small capacitance $C \to 0$. In this case, the correlation function in Eq.~(\ref{Definition of distribution function}) for the distribution function splits into two factors
\begin{equation}
\label{Correlation function. No fictitious contacts}
K^{\rm neq}_{\rm in,1}(t)=\sqrt{K^{\rm eq}_{\rm in}(t)} \cdot \sqrt{K^{\rm eq}_{\rm c}(t)}, 
\end{equation} 
where $K^{\rm eq}_{\alpha}(t)=-i \pi T_\alpha/\sinh[\pi T_{\alpha}(t-i0)]$, with $\alpha={\rm in,c}$. The square roots in Eq.~(\ref{Correlation function. No fictitious contacts}) arise from the fact that at $C \to 0$, the ROC acts as the current fractionalizer and $\mathcal{T}(\omega) \to 1/2$ in Eq.~(\ref{Solution to outgoing currents}). In case of cold ROC, $T_{\rm c}/T_{\rm in} \to 1$, expanding the integrand with respect to $T_{\rm c}/T_{\rm in}$ in Eq.~(\ref{Definition of distribution function}), the result for the distribution function to leading order is given by the equilibrium FD function at base temperature
\begin{equation}
\label{Distribution function. Cold contact. No fictitious Ohmic contact}
f^{\rm neq}_{\rm in,1}(\epsilon)=f_{\rm FD}(\epsilon/T_{\rm in}).	
\end{equation}
The sub-leading contribution is equal to $(1-T^2_{\rm c}/T^2_{\rm in})\partial f_{\rm FD}(z\epsilon)/4\partial z$ at $z \to 1$. Qualitatively, the result of Eq.~(\ref{Distribution function. Cold contact. No fictitious Ohmic contact}) is clear; the whole system is almost in equilibrium at $T_{\rm c}/T_{\rm in} \to 1$ and the result in leading order is given by the equilibrium distribution function of FD.  

In case of hot ROC, $T_{\rm c}/T_{\rm in} \to \infty$, one gets the following results in low and high energy regimes~\cite{supp} 
\begin{equation}
\label{Distribution function. Hot contact. No fictitious Ohmic contact}
\begin{aligned}
& f^{\rm neq}_{\rm in,1}(\epsilon) = 1/2-\mathcal{C} |\epsilon| /4 T_{\rm c}, \quad |\epsilon| /T_{\rm c} \ll 1, \\
& f^{\rm neq}_{\rm in,1}(\epsilon) = e^{-\epsilon/T_{\rm c}}/\sqrt{\pi \epsilon/T_{\rm c}}, \quad \epsilon/T_{\rm c} \gg 1,
\end{aligned}	
\end{equation}
where $\mathcal{C}=(2/\pi^2)\int dx \sqrt{x/\sinh(x)} \approx 1.5573$ and   $f^{\rm neq}_{\rm in}(\epsilon) \to 1$ at $\epsilon \to-\infty$. From Eq.~(\ref{Distribution function. Hot contact. No fictitious Ohmic contact}), one can see that the tail of the distribution function differs sufficiently from the equilibrium one at temperature $T_{\rm c}$, which has a pure exponential tail $\exp(-\epsilon/T_{\rm c})$. Moreover, the sub-leading term in the low energy regime is different by factor $\mathcal{C}>1$ compared to the equilibrium result $\mathcal{C}=1$. It is worth mentioning that in case of cold ROC the leading order contribution is given by the FD factor for both low- and high energy regimes. In contrast, the hot ROC's case indicates the non-equilibrium nature of the prepared state more precisely in both low- and high energy regimes, since the temperature difference between incoming and outgoing states is huge, $T_{\rm c}/T_{\rm in} \to \infty$. Despite the conventional assumption that the ROC completely equilibrates the incoming states, our results, Eqs.~(\ref{Zero frequency noise for outgoing state with N Ohmic contacts}) and (\ref{Distribution function. Hot contact. No fictitious Ohmic contact}), show that this picture is not accurate. These findings emphasize the limitations of assuming complete equilibration at the ROC.

In order to model the relaxation at the QH edge, we consider the case of one large FOC ($C_{\rm f} \to \infty$) and $T_{\rm in} R_{\rm q} C_{\rm f} \gg 1$. In this case according to the scattering problem in Eq.~(\ref{Outgoing current after Nth fictitious contact}), the correlation function in Eq.~(\ref{Definition of distribution function}) is given by the product of the equilibrium two-point correlation function and the non-equilibrium part~\cite{supp} 
\begin{equation}
\label{Correlation function. One large fictitious contact}
K_{\rm out, 1}(t)=K^{\rm eq}_{\rm f}(t) \cdot \exp\left[\Phi(t)\right],
\end{equation}
where $K^{\rm eq}_{\rm f}(t)$ is the equilibrium two-point correlation function at temperature $T_{\rm f}$ as in Eq.~(\ref{Correlation function. No fictitious contacts}) and the dimensionless function is responsible for non-equilibrium effects in exponent given by the integral over frequency~\cite{supp} 
\begin{equation}
\label{Nonequilibrium part of correlation function. One large fictitious Ohmic contact}	
\Phi(t)=R_{\rm q} \int d\omega \frac{e^{-i\omega t}-1}{\omega^2+\eta^2} [S_{\rm out,1}(\omega)-S^{\rm eq}_{\rm f}(\omega)], \quad \eta \to 0,
\end{equation}
and, to recall, $S_{\rm out, 1}(\omega)$ is given in Eq.~(\ref{Outgoing noise. Main text result}). In equilibrium $\exp[\Phi(t)]=1$ and $K^{\rm eq}_{\rm f}(t)=K^{\rm eq}_{\rm in}(t)$. Further, we substitute Eq.~(\ref{Correlation function. One large fictitious contact}) in Eq.~(\ref{Definition of distribution function}) and take into account that the main contribution to the integral comes from times $t \lesssim R_{\rm q}C_{\rm f}$, where one can replace the function $|\mathcal{A}(\omega)|^2= 1/[1+(\omega R_{\rm q} C_{\rm f})^2]$ in the integrand of $\Phi(t)$ in Eq.~(\ref{Nonequilibrium part of correlation function. One large fictitious Ohmic contact}) by the delta function, $\pi \delta(\omega)/R_{\rm q} C_{\rm f}$. After some algebra, one arrives at the following result~\cite{supp} 
\begin{equation}
\label{Distribution function with one big fictitious Ohmic contact}
\begin{aligned}
& f_{\rm out, 1}(\epsilon)=f_{\rm FD}(\epsilon/T_{\rm f}) + \delta f(\epsilon),\\
& \delta f(\epsilon)=\frac{\pi^2}{C_{\rm f}} \lim_{\omega \to 0}\left[S^{\rm neq}_{\rm in,1}(\omega)-S^{\rm eq}_{\rm f}(\omega)\right] \frac{\partial^2 }{\partial \epsilon^2} f_{\rm FD}(\epsilon/T_{\rm f}).
\end{aligned}
\end{equation}
The pre-factor in the sub-leading term is given by zero frequency, i.e. $\lim_{\omega \to 0}\left[S^{\rm neq}_{\rm in,1}(\omega)-S^{\rm eq}_{\rm f}(\omega)\right]=G_{\rm q}(T_{\rm in}+T_{\rm c}-2T_{\rm f})/2$.
From the above equation, one can see that at $T_{\rm f}=T_{\rm c}=T_{\rm in}$ the system is in equilibrium and the distribution function is given only by the leading order term, which indeed has the FD form at base temperature. 

%===============================================================================
\textit{Noise generated by QPC.}
The non-symmetrized zero-frequency noise can be calculated using the Landauer-Buttiker scattering approach locally at QPC [52]. The noise of the outgoing current in channel "$\rm out, 1$"
after QPC is found to be given by the sum of two terms
\begin{equation}
\label{Noise after QPC as sum of two terms}
S^{11}_{\rm out}(0)/G_{\rm q}=\mathcal{S}_{\rm t}(0) + \mathcal{S}_{\rm g}(0).
\end{equation}
The first term corresponds to the transmitted noise and
has the form
\begin{equation}
\label{Transmitted noise after QPC}
\mathcal{S}_{\rm t}(0) =\mathcal{D}^2T_{\rm in}+\frac{1}{2}\mathcal{R}^2(T_{\rm in}+T_{\rm c}),
\end{equation}
where $\mathcal{D}$ is the transparency of the QPC and $\mathcal{R}=1-\mathcal{D}$. The first term is the contribution of incoming channel 3 and the second term is responsible for the noise coming after the ROC in the regime of charge fractionalizer (see Fig.~\ref{fig:one}). One can see that in an almost equilibrium condition, $T_{\rm c} \to T_{\rm in}$, the transmitted noise is given by the equilibrium base temperature, $\mathcal{S}_{\rm t}(0)=T_{\rm in}$. The second term in Eq.~(\ref{Noise after QPC as sum of two terms})  describes the noise generated by QPC due to mixing of the incoming states from ROC and channel 3. It is given by the following time integral~\cite{supp} 
\begin{equation}
\label{cross term in total noise}
\mathcal{S}_{\rm g}(0)/2\mathcal{R}\mathcal{D}  \propto \int dt K^{\rm neq}_{\rm in,1}(t) \cdot K^{\rm eq}_{\rm in,3}(t),
\end{equation}
where the first factor is given in Eq.~(\ref{Correlation function. No fictitious contacts}) in the fractionalizer regime and the second multiplier corresponds to free fermions at temperature $T_{\rm in}$ originating from the channel ${\rm in,3}$", thus $K^{\rm eq}_{\rm in,3}(t)\propto -iT_{\rm in}/2\sinh(\pi T_{\rm in}t-i0)$.

For cold ROC, in the leading order of $T_{\rm c}/T_{\rm in} \to 1$, one can replace the contact's temperature $T_{\rm c}$ with $T_{\rm in}$ and obtain the cross-term~\cite{supp} 
\begin{equation}
\label{Cross noise at QPC Cold ROC}
\mathcal{S}_{\rm g}(0)/2\mathcal{R}\mathcal{D} = T_{\rm in}.
\end{equation}
The sub-leading correction to Eq.~(\ref{Cross noise at QPC Cold ROC}) is proportional to $T_{\rm c}-T_{\rm in} \ll T_{\rm in}$. In case of a hot ROC at $T_{\rm c}/T_{\rm in} \to \infty$, one can replace the finite-temperature $T_{\rm in}$ correlation functions in the integrand of Eq.~(\ref{cross term in total noise}) with the zero temperature case and obtain the leading order behavior as follows~\cite{supp}  
\begin{equation}
\label{Cross noise at QPC Hot ROC}
\mathcal{S}_{\rm g}(0)/2\mathcal{R}\mathcal{D}= \mathcal{M} T_{\rm c} , \quad \mathcal{M} \approx 0.4739.
\end{equation}
The sub-leading correction to Eq.~(\ref{Cross noise at QPC Hot ROC}) is proportional to $T_{\rm c}(T^2_{\rm in}/T^2_{\rm c}) \ll T_{\rm c}$.

In the case of FOC, the transmitted noise in Eq.~(\ref{Noise after QPC as sum of two terms}) is given by Eq.~(\ref{Transmitted noise after QPC}). The noise generated by QPC has the following integral form~\cite{supp} 
\begin{equation}
\label{Generated noise at QPC after the ROC}
\mathcal{S}_{\rm g}(0)/2\mathcal{R} \mathcal{D} \propto \int dt  K^{\rm eq}_{\rm{in,3}}(t) \cdot K^{\rm eq}_{\rm f}(t) \cdot\exp[\Phi(t)],
\end{equation}
where $\Phi(t)$ is given in Eq.~(\ref{Nonequilibrium part of correlation function. One large fictitious Ohmic contact}). The effect of $\Phi(t)$ is perturbative and the main contribution to the integral, as in the case of the distribution function above, comes from times $t \lesssim R_{\rm q} C_{\rm f}$. First, we consider the case of cold Ohmic contacts, namely $T_{\rm c} \to T_{\rm f} \to T_{\rm}$. In this case, one can replace all temperatures in leading order with $T_{\rm in}$ and get~\cite{supp} 
\begin{equation}
\label{Equation for cross term of noise for FOC at cold contacts case}
\begin{split}
& \mathcal{S}_{\rm g}(0)/2\mathcal{R} \mathcal{D} = T_{\rm in} +\delta T_{\rm g}(0), \\
& \delta T_{\rm g}(0) = \frac{\pi^2}{6T_{\rm in} C_{\rm f}} \lim_{\omega \to 0} [S^{\rm neq}_{\rm{in,1}}(\omega)-S^{\rm eq}_{\rm f}(\omega)].
\end{split}
\end{equation}

In this case of hot Ohmic contacts $T_{\rm c} \sim T_{\rm f} \gg T_{\rm in}$ one can replace the correlation function at temperature $T_{\rm in}$ in the leading order with zero temperature case and get~\cite{supp} 
\begin{equation}
\label{Equation for cross term of noise for FOC at hot contacts case}
\begin{split}
& \mathcal{S}_{\rm g}(0)/2\mathcal{R} \mathcal{D} = T_{\rm f} \ln{2} +\delta \tilde{T}_{\rm g}(0), \\
& \delta \tilde{T}_{\rm g}(0) = \frac{\pi^2}{4T_{\rm f} C_{\rm f}} \lim_{\omega \to 0} [S^{\rm neq}_{\rm{in,1}}(\omega)-S^{\rm eq}_{\rm f}(\omega)].
\end{split}
\end{equation}

In summary, we have investigated thermal relaxation at a quantum Hall edge, modeled by the FOC with a large capacitance. The initial nonequilibrium state is generated by the ROC in the charge-fractionalization regime, providing an alternative mechanism to QPC partitioning. Both, electron distribution function and the electric current noise are calculated. For a cold ROC, when the system is close to equilibrium and thermalization effects are negligible, the electron distribution function and the noise at the quantum Hall edge are described by the corresponding equilibrium expressions determined by the FD distribution. However, when thermalization effects become significant, the outgoing state immediately downstream of the ROC cannot be described by an equilibrium FD distribution and by Johnson–Nyquist noise at the ROC temperature, despite the conventional assumption of complete equilibration by the ROC. Furthermore, we calculated the electron distribution function and the current noise generated by a QPC placed downstream of the ROC and FOC. We found that the subleading corrections to both the electron distribution function and the QPC-generated noise contain a universal prefactor determined by the difference between the nonequilibrium noise generated by the ROC and the noise of a free chiral channel.   

%We are grateful to ... for fruitful discussion.
The author thanks Patrick R. Behr for careful reading of the manuscript and fruitful discussions. The author acknowledges financial support from the United Arab Emirates University under the Startup
Grant No. G00004974, UPAR Grant G00005479, and UAEU-AUA Grant G00005522.

\bibliography{refs}

\end{document}

\title{Thermal-energy relaxation in chiral quantum Hall edge mode \\ Supplemental material}

\author{Edvin G. Idrisov}
\affiliation{Department of Physics, United Arab Emirates University, P.O. Box 15551 Al-Ain, United Arab Emirates}

\date{\today}

% The following information is for internal review, please remove them for submission
%\widetext
%\leftline{Version xx as of \today}
%\leftline{Primary authors: Joe E. Physics}
%\leftline{To be submitted to (PRL, PRD-RC, PRD, PLB; choose one.)}
%\leftline{Comment to {\tt d0-run2eb-nnn@fnal.gov} by xxx, yyy}
%\centerline{\em D\O\ INTERNAL DOCUMENT -- NOT FOR PUBLIC DISTRIBUTION}

% the following line is for submission, including submission to the arXiv!!
%\hspace{5.2in} \mbox{Fermilab-Pub-04/xxx-E}

%\title{Template for PRL/PRD Papers}
%\input author_list.tex       % D0 authors (remove the first 3 lines
                             % of this file prior to submission, they
                             % contain a time stamp for the authorlist)
                             % (includes institutions and visitors)
%\date{\today}

%\pacs{}
\maketitle

%\section{\label{sec:level1}First-level heading}
% sections are not used for PRL papers

%==================================================================================
\section{A: Heat current in chiral edge mode}
\label{Sec:A}
Using the definition of heat flux in the main text, $J_{\alpha}=R_q \langle j^2_{\alpha}(t)\rangle/2$ and subtracting the contribution of the ground state, we get the following expression for the thermal current
\begin{equation}
\label{Heat current in chiral edge mode}
J_{\alpha}= \frac{R_q}{2} \int \frac{d\omega}{2\pi}\left[S^{\rm{eq}}_{\alpha}(\omega)-S_q(\omega)\right]=\int_0^{\infty} \frac{d\omega}{2\pi} \frac{\omega}{e^{\omega/T_{\alpha}}-1}=\frac{\pi T^2_{\alpha}}{12}, \quad \alpha=\rm{in, c, f}.
\end{equation}
This is the quantum heat flux and can be as well derived using FDT in equilibrium. If the channel is additionally biased, then according to Joules law, one needs to add the term proportional to square of electric current, namely $V^2/2R_q$. The Eq.~(\ref{Heat current in chiral edge mode}) will be used below to calculate the thermalization effects in real and fictitious Ohmic contacts.

\section{B: Non-equilibrium finite frequency noise for real Ohmic contact}
\label{Sec:B}
We consider the real Ohmic contact (ROC) with capacitance $C$ as a source of heat current here. The solution to the outgoing current in channel "1" in the frequency representation for the case of ROC is expressed in terms of the incoming currents and the Langevin sources
\begin{equation}
\label{Solution to outgoing current for real Ohmic contact}
j^{\rm neq}_1(\omega)=j_{{\rm c},1}(\omega)+\sum_{l=1,2} \mathcal{T}(\omega)\left[j_{{\rm in},l}(\omega)-j_{{\rm c},l}(\omega)\right], \quad \mathcal{T}(\omega)=1/(2-i\omega R_{\rm q} C).
\end{equation}
The first term of the scattering coefficient $\mathcal{T}(\omega)$ in the denominator counts the number of incoming and outgoing channels with respect to the ROC. Next, using Eq.~(\ref{Solution to outgoing current for real Ohmic contact}), one can calculate the finite frequency noise of the outgoing current right after the ROC, namely $S^{\rm neq}_1(\omega)=\int dt e^{i\omega t} \langle \delta j^{\rm neq}_{1}(t) \delta j^{\rm neq}_{1}(0) \rangle$ and $\delta j^{\rm neq}_{1}(t)=j^{\rm neq}_{1}(t)-\langle j^{\rm neq}_{1}(t)\rangle$. After some algebra, the result is given by the following expression
\begin{equation}
\label{Non-equilibrium finite frequency noise for real Ohmic contact}
S^{\rm neq}_1(\omega)/G_{\rm q}=2|\mathcal{T}(\omega)|^2 \left[\frac{\omega}{1-e^{-\omega/T_{\rm in}}}+\frac{\omega}{1-e^{-\omega/T_{\rm c}}}\right]+[1-2{\rm Re}\mathcal{T}(\omega)] \frac{\omega}{1-e^{-\omega/T_{\rm c}}},
\end{equation} 
where the ROC temperature, $T_{\rm c}$ is unknown and can be calculated using self-consistent energy balance equations, i.e., the conservation of the incoming and outgoing heat currents. When real Ohmic contact functions as fractionalizer, the scattering coefficient is $\mathcal{T}(\omega) \to 1/2$ at $C \to 0$. Then the finite frequency noise is equal to
\begin{equation}
\label{Non-equilibrium finite frequency noise for real Ohmic contact in fractionalizer regime}
S^{\rm neq}_1(\omega)/G_{\rm q}=\frac{1}{2} \left[\frac{\omega}{1-e^{-\omega/T_{\rm in}}}+\frac{\omega}{1-e^{-\omega/T_{\rm c}}}\right]=\frac{R_{\rm q}}{2}\left[S^{\rm eq}_{\rm in}(\omega)+S^{\rm eq}_{\rm c}(\omega)\right].
\end{equation} 
Thus, two incoming equilibrium noises are split into two parts due to two outgoing chiral channels from ROC. This result is given in Eq.~(11) of the main text. We are not interested in the noise after the fictitious Ohmic contact (FOC), but similarly one can show that after the FOC, i.e. $S_{\rm{out, 1}}(\omega) = |\mathcal{A}(\omega)|^2 \left[S^{\rm{neq}}_1(\omega)+S^{\rm{eq}}_{\rm f}(\omega)\right]+\left[1-2{\rm Re}\mathcal{A}(\omega)\right]S^{\rm eq}_{\rm f}(\omega)$, where $\mathcal{A}(\omega)=1/(1-i\omega R_{\rm q} C_{\rm f})$ and $S^{\rm eq}_{\rm f}(\omega)$ is an equilibrium spectral function at FOC's temperature $T_{\rm f}$.

\section{C: Determination of temperature $T_{\rm c}$ of real Ohmic contact}
\label{Sec:C}
To find the temperature $T_{\rm c}$, we solve the equations of the energy balance
self-consistently, i.e., conservation of the incoming and outgoing
heat currents, where the heat current is defined by Joule’s law, 
\begin{equation}
\label{Heat balance equation to find Tc}
(R_{\rm q}/2)\sum_{i=1,2} [\langle \delta j_{{\rm in},i}(t) \delta j_{{\rm in},i}(t)\rangle + \langle j_{{\rm in},i}(t)\rangle^2]
=
(R_{\rm q}/2)\sum_{i=1,2} [\langle \delta j^{\rm neq}_i(t) \delta j^{\rm neq}_i(t)\rangle + \langle j^{\rm neq}_i(t)\rangle^2],
\end{equation}
where $j^{\rm neq}_i(t)$ outgoing non-equilibrium current emitted from ROC. Using the above equation and the scattering matrix of currents for ROC, Eq.~(\ref{Solution to outgoing current for real Ohmic contact}), we arrive at the following expression for determining contact's temperature.
\begin{equation}
\label{General equation for determination of temperature ROC}
\begin{split}
& \frac{V^2_1}{2R_{\rm q}}+\frac{V^2_1}{2R_{\rm q}}+2 \frac{R_{\rm q}}{2} \int \frac{d\omega}{2\pi} \left[S^{\rm eq}_{\rm in}(\omega)-S_{\rm q}(\omega)\right]-2 \frac{R_{\rm q}}{2} \int \frac{d\omega}{2\pi}2|\mathcal{T}(\omega)|^2 \left[S^{\rm eq}_{\rm in}(\omega)-S_{\rm q}(\omega)\right] \\
& =2 \frac{R_{\rm q}}{2}\int \frac{d\omega}{2\pi}\left[S^{\rm eq}_{\rm c}(\omega)-S_{\rm q}(\omega)\right]- 2 \frac{R_{\rm q}}{2} \int \frac{d\omega}{2\pi}2|\mathcal{T}(\omega)|^2 \left[S^{\rm eq}_{\rm c}(\omega)-S_{\rm q}(\omega)\right]. 
\end{split}
\end{equation}
The first integrals on the left hand and right hand sides are given by Eq.~(\ref{Heat current in chiral edge mode}). 
%\begin{equation}
%\label{Final integral expression for the determination of real Ohmic contact's temperature}
%\begin{split}
%& \frac{V^2_1+V^2_2}{2R_{\rm q}}+2\underbrace{\int \frac{d\omega}{4\pi}\left[\frac{\omega }{1-%e^{-\omega/T_{\rm in}}}- \omega \theta(\omega)\right]}_{\pi T^2_{\rm in}/12}=2\int %\frac{d\omega}{4\pi}\frac{2}{(\omega R_{\rm q}C)^2+4}\left[\frac{\omega }{1-e^{-\omega/T_{\rm %in}}}- \omega \theta(\omega)\right]+\\
%& 2\underbrace{\int \frac{d\omega}{4\pi}\left[\frac{\omega }{1-e^{-\omega/T_{\rm c}}}- \omega %\theta(\omega)\right]}_{\pi T^2_{\rm c}/12}-2\int \frac{d\omega}{4\pi}\frac{2}{(\omega R_{\rm %q}C)^2+4}\left[\frac{\omega }{1-e^{-\omega/T_{\rm c}}}- \omega \theta(\omega)\right].
%\end{split}
%\end{equation}
The second integral with capacitance can be rewritten using special functions as follows ($\alpha=\rm{in, c}$)
\begin{equation}
\begin{split}
& - 2 \frac{R_{\rm q}}{2} \int \frac{d\omega}{2\pi}2|\mathcal{T}(\omega)|^2 \left[S^{\rm eq}_{\alpha}(\omega)-S_{\rm q}(\omega)\right]=-2\int \frac{d\omega}{4\pi}\frac{2}{(\omega R_{\rm q}C)^2+4}\left[\frac{\omega }{1-e^{-\omega/T_{\alpha}}}- \omega \theta(\omega)\right] \\
& =2\int_0^{\infty} \frac{d\omega}{4\pi} \frac{1}{1+(\omega R_{\rm q} C/2)^2}\frac{\omega}{1-e^{\omega/T_{\alpha}}}=\frac{\pi T^2_{\alpha}}{12} \cdot \mathcal{F}(\pi T_{\alpha} R_{\rm q} C/2), 
\end{split}
\end{equation}
where the dimensionless function equals the following expression
\begin{equation}
\label{Dimensionless integral}
\mathcal{F}(x)=\frac{3}{x^2}\int_0^{\infty} \frac{dz}{1+z^2} \frac{2z}{1-e^{\pi z/x}}=\frac{3}{x^2}\left[x+\log(2x)+\Psi(1/2x)\right],
\end{equation}
and $\Psi(x)=d \log \Gamma(x)/dx$ is a digamma function, and $\Gamma(x)=\int^{\infty}_0 y^{x-1}e^{-y}dy$ is a gamma function. Further, choosing the symmetric bias and thus setting $V_1=-V_2=V$, we obtain the equation to determine the temperature of ROC
\begin{equation}
\label{Final equation for determination of temperature of real Ohmic contact}	
\frac{V^2}{R_{\rm q}}+\frac{\pi T^2_{\rm in}}{12}\left[2+\mathcal{F}(\pi T_{\rm in} R_{\rm q} C/2)\right]=\frac{\pi T^2_{\rm c}}{12}\left[2+\mathcal{F}(\pi T_{\rm c} R_{\rm q} C/2)\right].
\end{equation}
We are interested when ROC functions as a fractionalizer, i.e., $C \to 0$. In this case $\mathcal{F}(x) \to -1$ at $x \to 0$ and we get the final result $T^2_{\rm c}=T^2_{\rm in}+6V^2/\pi^2$ from Eq.~(\ref{Final equation for determination of temperature of real Ohmic contact}), which is presented in Eq.~(8) of the main text. At $V=0$ the system is in equilibrium and at $T_{\rm c}=T_{\rm in}$.

\section{D: Determination of temperature $T_{\rm f}$ of fictitious Ohmic contact}
\label{Sec:D}  
 In order to calculate the distribution function and noise further, one needs to take into account the thermalization effects inside the fictitious Ohmic contact (FOC). For the FOC it is given by conservation law, namely the incoming heat flux is equal to the outgoing heat flux. Using the definition of heat flux and the same steps as in the previous section, we arrive at the following equation
\begin{equation}
\label{Equation for determination of temperature of fictitious Ohmic contact}
 \frac{R_q}{2} \int \frac{d\omega}{2\pi}(S^{{\rm neq}}_{{1}}(\omega)-S_{\rm q}(\omega))=\frac{R_q}{2} \int  \frac{d\omega}{2\pi} |\mathcal{A}(\omega)|^2(S^{{\rm neq}}_{{1}}(\omega)-S_{\rm q}(\omega))+\frac{R_q}{2} \int  \frac{d\omega}{2\pi} \left[1-|\mathcal{A}(\omega)|^2\right](S^{\rm eq}_{{\rm f,1}}(\omega)-S_{\rm q}(\omega))
\end{equation}
where $S^{\rm neq}_1(\omega)$ are given by Eq.~(\ref{Non-equilibrium finite frequency noise for real Ohmic contact}) for ROC. Next, we find the temperature $T_{\rm f}$ in the case of small and large capacitance $C_{\rm f}$. To find it, we can use the above equation Eq.~(\ref{Equation for determination of temperature of fictitious Ohmic contact})
\begin{equation}
\label{Second equation in this section}
\frac{R_{\rm q}}{2}\int \frac{d\omega}{2\pi} \left[1-|\mathcal{A}(\omega)|^2 \right] \left[S^{\rm neq}_{\rm in,1}(\omega)-S_{\rm q}(\omega)\right]=\frac{R_{\rm q}}{2}\int \frac{d\omega}{2\pi} \left[1-|\mathcal{A}(\omega)|^2 \right]  \left[S^{\rm eq}_{\rm f}(\omega)-S_{\rm q}(\omega)\right].
\end{equation}
Using Eq.~(\ref{Heat current in chiral edge mode}), the right hand side can be rewritten as follows
\begin{equation}
\label{General expression for left hand side to determine the temperature Tf}	
\frac{R_{\rm q}}{2}\int \frac{d\omega}{2\pi} \left[1-|\mathcal{A}(\omega)|^2 \right] \left[S^{\rm eq}_{\rm f}(\omega)-S_{\rm q}(\omega)\right]=\frac{\pi T^2_{\rm f}}{12}+\frac{R_{\rm q}}{2}\int \frac{d\omega}{2\pi} (-|\mathcal{A}(\omega)|^2) \left[S^{\rm eq}_{\rm f}(\omega)-S_{\rm q}(\omega)\right].
\end{equation}
After introducing the dimensionless variable $z=\omega R_q C_{\rm f}$, the second term of the above equation takes the form
\begin{equation}
\label{Integral I}	
\frac{R_{\rm q}}{2}\int \frac{d\omega}{2\pi} (-|\mathcal{A}(\omega)|^2) \left[S^{\rm eq}_{\rm f}(\omega)-S_{\rm q}(\omega)\right]= \frac{\pi T^2_{\rm f}}{12} \cdot \frac{3}{(\pi T_{\rm f} R_{\rm q} C_{\rm f})^2} \int_0^{\infty} \frac{dz }{1+z^2}\frac{2z}{1-e^{z/T_{\rm f} R_{\rm q} C_{\rm f}}} = \frac{\pi T^2_{\rm f}}{12} \cdot \mathcal{F}(\pi T_{\rm f} R_{\rm q} C_{\rm f}),	
\end{equation}
where $\mathcal{F}(x)$ is given in Eq.~(\ref{Dimensionless integral}). Thus, we obtain the following equation to determine the temperature of FOC
\begin{equation}
\label{Final equation for determination of temperature of fictitious Ohmic contacts}
\frac{R_{\rm q}}{2}\int \frac{d\omega}{2\pi} \left[1-|\mathcal{A}(\omega)|^2\right] \left[S^{\rm neq}_1(\omega)-S_{\rm q}(\omega)\right]=\frac{\pi T^2_{\rm f}}{12} \cdot \left[1+\mathcal{F}(\pi T_{\rm f} R_{\rm q} C_{\rm f})\right].
\end{equation}
For the case of ROC as a non-equilibrium source using in the fractionalizer regime, Eq.~(\ref{Non-equilibrium finite frequency noise for real Ohmic contact in fractionalizer regime}) we get the following expression for the left hand side of Eq.~(\ref{Second equation in this section})
\begin{equation}
\label{left hand side for temperature determination for real fractionalizer}
\frac{R_{\rm q}}{2}\int \frac{d\omega}{2\pi} \left[1-|\mathcal{A}(\omega)|^2 \right] \left[S^{\rm neq}_1(\omega)-S_{\rm q}(\omega)\right]=\frac{1}{2} \cdot \frac{\pi T^2_{\rm in}}{12}\cdot \left[1+\mathcal{F}(\pi T_{\rm in} R_{\rm q} C_{\rm f})\right]+\frac{1}{2} \cdot \frac{\pi T^2_{\rm c}}{12}\cdot \left[1+\mathcal{F}(\pi T_{\rm c} R_{\rm q} C_{\rm f})\right].
\end{equation}
Therefore, the final equation for the determination of the temperature of FOC takes the following form
\begin{equation}
\label{Equation for determination of temperature TF in case of real fractionalizer}
\frac{1}{2} \cdot \frac{\pi T^2_{\rm in}}{12}\cdot \left[1+\mathcal{F}(\pi T_{\rm in} R_{\rm q} C_{\rm f})\right]+\frac{1}{2} \cdot \frac{\pi T^2_{\rm c}}{12}\cdot \left[1+\mathcal{F}(\pi T_{\rm c} R_{\rm q} C_{\rm f})\right]=\frac{\pi T^2_{\rm f}}{12} \cdot \left[1+\mathcal{F}(\pi T_{\rm f} R_{\rm q} C_{\rm f})\right].
\end{equation}
At large capacitance $C_{\rm f} \to \infty$, one can use $\mathcal{F}(x) \approx -3/x$ for $x \gg 1$, thus the temperature of the FOC is given by $T^2_{\rm f}=\left(T^2_{\rm in}+T^2_{\rm c}\right)/2$. At small capacitance $C_{\rm f} \to 0$, one can use $\mathcal{F}(x) \approx -1+2x^2/5$ for $x \ll 1$ and the temperature of the FOC is given by $T^4_{\rm f}=\left(T^4_{\rm in}+T^4_{\rm c}\right)/2$. Our interest is the large capacitance case, which is discussed in the main text, Eq.~(9).

\section{E: Calculation of initial distribution function for hot ROC}
\label{Sec:E}
According to Eqs.~(2) and (9) of the main text, the initial non-equilibrium distribution function after ROC has the following form
\begin{equation}
\label{Initial distribution function for real Ohmic contact}
f^{\rm neq}_{\rm in,1}(\epsilon)=\int \frac{dt}{2\pi i}e^{-i\epsilon t}\sqrt{\frac{\pi T_{\rm in}}{\sinh(\pi T_{\rm in}t-i0)}}\sqrt{\frac{\pi T_{\rm c}}{\sinh(\pi T_{\rm c}t-i0)}},
\end{equation}
where $T_{\rm c}$ is given in Eq.~(\ref{Final equation for determination of temperature of real Ohmic contact}). Calculations for the case of "cold" ROC, $T_{\rm c}/T_{\rm in} \to 1$, are straightforward, and since the system is almost in equilibrium, the main contribution is given by equilibrium FD distribution function (See Eq.~(14) in the main text). In the regime of hot ROC, $T_{\rm c}/T_{\rm in} \to \infty$ and low energies $\epsilon /T_{\rm c} \ll 1$ (for definiteness, positive values of $\epsilon >0$ are considered), we introduce the dimensionless variable $x=\pi T_c t$, expand the oscillatory exponent and replace the correlation function with $T_{\rm in}$ by the zero temperature case, namely  
\begin{equation}
\label{Distribution function at low energies for hot Ohmic contact: expansion}
f^{\rm neq}_{\rm in,1}(\epsilon)=\int \frac{dx}{2\pi i} \frac{e^{-i\frac{\epsilon}{\pi T_{\rm c}}x}}{\sqrt{\sinh[x-i0]}} \frac{\sqrt{T_{\rm in}/T_{\rm c}}}{\sqrt{\sinh[(T_{\rm in}/T_{\rm c})(x-i0)]}}\approx \int \frac{dx}{2\pi i} \frac{1+\frac{-i\epsilon}{\pi T_{\rm c}}x}{\sqrt{x-i0}\sqrt{\sinh[x-i0]}}.
\end{equation}
Next, keeping the leading order terms, we get the result for initial non-equilibrium distribution function at low energies
\begin{equation}
\label{Hot contact approximate distribution function at low energies}
f^{\rm neq}_{\rm in,1}(\epsilon)=\frac{1}{2}-\frac{\mathcal{C} \epsilon}{4T_{\rm c}}, \quad \mathcal{C}=\frac{4}{\pi^2}\int_0^{\infty} \frac{\sqrt{x} dx}{\sqrt{\sinh x}} \approx 1.5573 >1.	
\end{equation} 
This result is given in Eq.~(15) of the main text. Now, for the case of high energies, $\epsilon/T_{\rm c} \gg 1$ the oscillatory term is essential and cannot be expanded in Eq.~(\ref{Distribution function at low energies for hot Ohmic contact: expansion}). Therefore, in the leading order of $T_{\rm c}/T_{\rm in} \to \infty$ the form of the integral is
\begin{equation}	
f^{\rm neq}_{\rm in,1}(\epsilon) \approx \int \frac{dx}{2\pi i} \frac{e^{-\frac{i\epsilon}{\pi T_c}x}}{\sqrt{\sinh(x-i0)}}\frac{1}{\sqrt{x-i0}}.
\end{equation} 
To estimate this integral, we first shift the integration variable $x=u-i\pi/2$ and get 
\begin{equation}
f^{\rm neq}_{\rm in,1}(\epsilon)=e^{-\frac{\epsilon}{2T_c}} \int \frac{du}{2\pi} \frac{e^{-\frac{i \epsilon}{\pi T_c} u}}{\sqrt{\cosh u}} \frac{1}{\sqrt{\frac{\pi}{2}+iu}}=\int \frac{dz}{\sqrt{\pi}} e^{-\frac{\pi z^2}{2}} \int \frac{du}{2\pi} \frac{e^{-i\left( \frac{\epsilon}{\pi T_c}+z^2 \right)u}}{\sqrt{\cosh u}}.
\end{equation} 
In the above equation, the inner integration with respect to the variable $u$ is given by a gamma function and has the form
\begin{equation}
f^{\rm neq}_{\rm in,1}(\epsilon)= \frac{1}{2\pi} e^{-\frac{\epsilon}{2T_c}}\int \frac{dz}{\sqrt{\pi}} e^{-\frac{\pi z^2}{2}} \frac{1}{\sqrt{2\pi}} \left|\Gamma\left[1/4+i\left(\epsilon/\pi T_c+z^2\right)/2\right]\right|^2.
\end{equation} 
Using further $|\Gamma(x+iy)|^2 \simeq 2\pi y^{2x-1} e^{-\pi y}$ at $y \to +\infty$ we arrive at the following result
\begin{equation}
f^{\rm neq}_{\rm in,1}(\epsilon) = \frac{1}{\pi} e^{-\frac{\epsilon}{T_c}} \int \frac{e^{-\pi z^2} dz}{\sqrt{z^2+\epsilon/\pi T_c}}=\frac{1}{\pi} e^{-\frac{\epsilon}{2T_c}} \cdot K_0(\epsilon/2T_c),
\end{equation}
where $K_0(x)$ is the modified Bessel function of the second kind. Taking into account that $K_0(x) \simeq \sqrt{\pi/2x}e^{-x}$ for $x \gg 1$ we arrive at Eq.~(15) from the main text.

\section{F: Calculation of distribution function after FOC}
\label{Sec:F}
The distribution function after the FOC can be written as follows
\begin{equation}
\label{Distribution function through second and third cummulants}
f_{\rm out, 1}(\epsilon)=\int \frac{dt}{2\pi} e^{-i(\epsilon-2\pi I_{\rm in})t} \cdot K^{\rm eq}_{\rm f}(t) \cdot e^{\Phi(t)}, 
\end{equation}
where $I_{\rm in}=\langle j^{\rm neq}_1\rangle=0$ is an incoming non-equilibrium current after ROC and is equal to zero since the real Ohmic contact is not biased $V_1+V_2=0$, and $K^{\rm eq}_{\rm f}(t) \propto -iT_{\rm f}/\sinh(\pi T_{\rm f}t-i0)$ is a two-point free-fermion correlation function. The second factor is given by 
\begin{equation}
\label{Equation for Phi(t)}
\Phi(t)= R_{\rm q} \int d\omega \frac{e^{-i\omega t}-1}{\omega^2+\eta^2} \frac{S^{\rm neq}_{1}(\omega)-S^{\rm eq}_{\rm f}(\omega)}{1+(\omega R_{\rm q} C_{\rm f})^2}, \quad \eta \to +0,
\end{equation}
One can see that $\Phi^{\ast}(t)=\Phi(t)=\Phi(-t)$ and note that at $T_{\rm in}=T_{\rm c}=T_{\rm f}$, $\Phi(t)=0$, and we recover the equilibrium Fermi-Dirac distribution.We are interested in large capacitance FOC at $C_{\rm f} \to \infty$, thus the effect of $\Phi(t)$ is perturbative and expanding the exponent in Eq.~(\ref{Distribution function through second and third cummulants}) we have 
\begin{equation}
\label{Distribution function as sum of f1, f2, f3}	
f_{\rm out, 1}(\epsilon) \approx f_1(\epsilon)+f_2(\epsilon).
\end{equation}
The first term in Eq.~(\ref{Distribution function as sum of f1, f2, f3}) is given by the equilibrium FD distribution function at temperature $T_{\rm f}$
\begin{equation}
\label{f1 term}	
f_1(\epsilon)=\int \frac{dt}{2\pi} e^{-i\epsilon t} \cdot K^{\rm eq}_{\rm f}(t)=f_{\rm FD}(\epsilon/T_{\rm f})=\frac{1}{1+e^{\epsilon/T_{\rm f}}}. 
\end{equation}
The second term is given by 
\begin{equation}
\label{f2 term}	
f_2(\epsilon)= R_{\rm q}\int \frac{dt}{2\pi} e^{-i\epsilon t} \cdot K^{\rm eq}_{\rm f}(t) \cdot \Phi(t) = \int \frac{dt}{2\pi} e^{-i \epsilon t} \cdot K^{\rm eq}_{\rm f}(t)\int d\omega \frac{e^{-i\omega t}-1}{\omega^2+\eta^2} \frac{S^{\rm neq}_1(\omega)-S^{\rm eq}_{\rm f}(\omega)}{1+(\omega R_{\rm q} C_{\rm f})^2}.
\end{equation}
At large capacitance $C_{\rm f} \to \infty$, the main contribution to the integral comes from finite times $t \lesssim R_{\rm q} C_{\rm f}$ and small frequencies $\omega t \ll 1$, where the fraction $1/(1+(\omega R_{\rm q} C_{\rm f})^2)$ can be replaced by the delta function  $(\pi/R_{\rm q} C_{\rm f})\delta(\omega)$. Therefore, we get
\begin{equation}
f_2(\epsilon)=\frac{-\pi^2}{C_{\rm f}}\lim_{\omega \to 0}\left[S^{\rm neq}_1(\omega)-S^{\rm eq}_{\rm f}(\omega)\right] \int \frac{dt}{2\pi} e^{-i \epsilon t} t^2K^{\rm eq}_{\rm f}(t)  = \frac{\pi^2}{C_{\rm f}} \lim_{\omega \to 0}\left[S^{\rm neq}_1(\omega)-S^{\rm eq}_{\rm f}(\omega)\right] \cdot \frac{\partial^2 }{\partial \epsilon^2} f_{\rm FD}(\epsilon/T_{\rm f}).
\end{equation}
This result is given in Eq.~(18) of the main text. The prefactor can be estimated using Eq.~(\ref{Non-equilibrium finite frequency noise for real Ohmic contact}) for ROC. That is, 
\begin{equation}
\lim_{\omega \to 0}\left[S^{\rm neq}_1(\omega)-S^{\rm eq}_{\rm f}(\omega)\right]=(G_{\rm q}/2)(T_{\rm in}+T_{\rm c}-2T_{\rm f}).
\end{equation}
The above two results are presented in Eq.~(18) of the main text.

\section{G: QPC-generated noise right after the ROC}
\label{Sec: G}
Using Landauer-Buttiker scattering approach, the non-symmetrized zero frequency noise (at zero frequency the non-symmetrized and symmetrized noises coincide) in the channel $1$ right after the QPC is given by
\begin{equation}
\label{Noise}
S^{11}_{\rm out}(\omega=0)/G_{\rm q}=\mathcal{D}^2\mathcal{P}_{33}+\mathcal{R}^2\mathcal{P}_{11}(\omega=0)+2\mathcal{R}\mathcal{D}\mathcal{P}_{13}(\omega=0).
\end{equation} 
In the above Eq.~(\ref{Noise}) the first and second terms are given by the FDT theorem; therefore,
\begin{equation}
\label{Diract terms for noise in case of ROC}
\mathcal{P}_{33}(\omega=0) \propto \int dt \langle \delta j_{\rm in,3}(t) \delta j_{\rm in,3}(0)\rangle=T_{\rm in}, \quad \mathcal{P}_{11}(\omega=0) \propto \int dt \langle \delta j^{\rm neq}_1(t) \delta j^{\rm neq}_1(0) \rangle= (T_{\rm in} + T_{\rm c})/2.
\end{equation}
The last term of Eq.~(\ref{Noise}) is given by the following time integral 
\begin{equation}
\label{cross term in total noise}
\mathcal{P}_{13}(\omega=0) \propto\int dt \langle (\psi^{\rm neq}_1)^{\dagger}(t) \psi^{\rm neq}_1(0)\rangle \langle \psi_{\rm in,3}(t) \psi^{\dagger}_{\rm in,3}(0)\rangle=\int dt K^{\rm neq}_{\rm in,1}(t) \cdot K^{\rm eq}_{\rm in,3}(t).
\end{equation}
Here, the second multiplier corresponds to free fermions at temperature $T_{\rm in}$ originating from the channel "${\rm in,3}$", thus $K^{\rm eq}_{\rm in,3}(t)\propto -iT_{\rm in}/2\sinh(\pi T_{\rm in}t-i0)$.
The first multiplier can be presented as follows
\begin{equation}
K^{\rm neq}_1(t) \propto \langle e^{2\pi i\int_{-\infty}^t j^{\rm neq}_1(t^{\prime})e^{\eta t^{\prime}}dt^{\prime}}e^{-2\pi i \int_{-\infty}^0 j^{\rm neq}_1(t^{\prime})e^{\eta t^{\prime}} dt^{\prime}}\rangle=\exp[M^{\rm neq}_1(t)], 
\end{equation}
After some algebra, one can rewrite it as follows 
\begin{equation}
M^{\rm neq}_1(t)= i R_{\rm q} \int^t_0 dt^{\prime} \langle j^{\rm neq}_1(t^{\prime})\rangle+ R_{\rm q} \int d\omega \frac{e^{-i\omega t}-1}{\omega^2+\eta^2} S^{\rm{neq}}_1(\omega), \quad \eta \to +0.
\end{equation}
Here, the first term is zero because the outgoing state of the ROC is not biased; therefore, we have  
\begin{equation}
\label{Exponent of correlation function}
M^{\rm neq}_1(t)=R_{\rm q} \int d\omega \frac{e^{-i\omega t}-1}{\omega^2+\eta^2}S^{\rm{neq}}_{\rm{in,1}}(\omega),
\end{equation}
where $S^{\rm neq}_1(\omega)$ is given in Eq.~(\ref{Non-equilibrium finite frequency noise for real Ohmic contact in fractionalizer regime}). 
We are interested in the limit of $C \to 0$ (fractionalizer); thus, using Eq.~(\ref{Non-equilibrium finite frequency noise for real Ohmic contact in fractionalizer regime}) one gets
\begin{equation}
M^{\rm neq}_1(t)=\int d\omega \frac{e^{-i\omega t}-1}{\omega^2+\eta^2} \left[\frac{\omega/2}{1-e^{-\omega/T_{\rm in}}}+\frac{\omega/2}{1-e^{-\omega/T_{\rm c}}} \right]=\log\left[\sqrt{\frac{-iT_{\rm in}}{2\sinh(\pi T_{\rm in}t-i0)}}\sqrt{\frac{-iT_{\rm c}}{2\sinh(\pi T_{\rm c}t-i0)}}\right].
\end{equation}
After substitution into Eq.~(\ref{cross term in total noise}), we finally get the following integral
\begin{equation}
\label{Final integral equation for cross noise term for ROC}
\mathcal{P}_{13}(\omega=0)=\int \frac{dt}{2\pi} \sqrt{\frac{-i\pi T_{\rm in}}{\sinh(\pi T_{\rm in}t-i0)}} \sqrt{\frac{-i\pi T_{\rm c}}{\sinh(\pi T_{\rm c}t-i0)}} \frac{-i\pi T_{\rm in}}{\sinh(\pi T_{\rm in}t-i0)}. 
\end{equation}
For cold ROC in the leading order of $T_{\rm c}/T_{\rm in} \to 1$, one can replace the contact's temperature $T_{\rm c}$ with $T_{\rm in}$, and introducing the dimensionless variable $x=\pi T_{in} t$ we obtain the following. 
\begin{equation}
\mathcal{P}_{13}(\omega=0) = T_{\rm in} \int \frac{(-1/2)dx}{\sinh^2(x-i0)}=T_{\rm in}.
\end{equation}
The sub-leading correction is proportional to $T_{\rm c}-T_{\rm in} \ll T_{\rm in}$. In case of hot ROC at $T_{\rm c}/T_{\rm in} \to \infty$, one can replace the finite-temperature $T_{\rm in}$ correlation functions in the integrand of Eq.~(\ref{Final integral equation for cross noise term for ROC}) with zero temperature case and introducing the dimensionless variable $x=\pi T_{\rm c} t$ we obtain the leading order behavior  
\begin{equation}
\mathcal{P}_{13}(0)= \mathcal{M} T_{\rm c} , \quad \mathcal{M} = \int \frac{(-1/2)dx}{(x-i0)^{3/2}\sqrt{\sinh(x-i0)}}\approx 0.4739.
\end{equation}
The sub-leading correction is proportional to $T_{\rm c}(T^2_{\rm in}/T^2_{\rm c}) \ll T_{\rm c}$.

\section{H: QPC-generated noise right after the FOC}
\label{Sec: H}
In case of FOC, the first two terms $\mathcal{P}_{11}(\omega=0)$ and $\mathcal{P}_{33}(\omega=0)$ are given by Eq.~(\ref{Diract terms for noise in case of ROC}). The last term of noise has the following integral form 
\begin{equation}
\label{P13 after QPC}
\mathcal{P}_{13}(\omega=0) \propto \int dt  K^{\rm eq}_{\rm{in,3}}(t) \cdot K^{\rm eq}_{\rm f}(t) \cdot e^{\Phi(t)}=\int \frac{dt}{2\pi} \frac{-i \pi T_{\rm in}}{\sinh(\pi T_{\rm in}t-i0)} \times \frac{-i\pi T_{\rm f}}{\sinh(\pi T_{\rm f}t-i0)} \times e^{\Phi(t)}
\end{equation}
where $\Phi(t)$ is given in Eq.~(\ref{Equation for Phi(t)}). At large frequencies of FOC the effect of $\Phi(t)$ is perturbative; therefore, the leading two terms are given by $\mathcal{P}_{13}(\omega =0)= \mathcal{I}_1+\mathcal{I}_2$ and 
\begin{equation}
\label{Equation for cross term of noise for FOC}
\mathcal{I}_1=\int \frac{dt}{2\pi} \frac{-i \pi T_{\rm in}}{\sinh(\pi T_{\rm in}t-i0)}\frac{-i\pi T_{\rm f}}{\sinh(\pi T_{\rm f}t-i0)}, \quad \mathcal{I}_2=\int \frac{dt}{2\pi} \frac{-i \pi T_{\rm in}}{\sinh(\pi T_{\rm in}t-i0)}\frac{-i\pi T_{\rm f}}{\sinh(\pi T_{\rm f}t-i0)} \times \Phi(t).
\end{equation}
First, we consider the case of cold Ohmic contacts, namely $T_{\rm c} \sim T_{\rm f} \to T_{\rm}$. In this case, we can replace $T_{\rm f}$ in the first term $\mathcal{I}_1$ with $T_{\rm in}$ and introducing the dimensionless variable of integration $x=\pi T_{\rm{in}}$ we obtain the following.  
\begin{equation}
\label{Integral I1 in case of cold ohmic contacts}
\mathcal{I}_1 = T_{\rm in}\int \frac{(-1/2)dx}{\sinh^2(x-i0)} = T_{\rm in}.
\end{equation}
The sub-leading correction is proportional to $T_{\rm f}-T_{\rm in} \ll T_{\rm in}$. The second integral can be estimated as in Sec.~F and after some algebra we obtain the following result,
\begin{equation}
\label{Integral I2 in case of cold ohmic contacts}
\mathcal{I}_2=\frac{1}{T_{\rm in} C_{\rm f}} \lim_{\omega \to 0} [S^{\rm neq}_1(\omega)-S^{\rm eq}_{\rm f}(\omega)] \int \frac{x^2 dx}{2\sinh^2(x-i0)} = \frac{\pi^2}{12 T_{\rm in} R_{\rm q} C_{\rm f}}(T_{\rm c}+T_{\rm in}-2T_{\rm f}).
\end{equation}
In this case of hot Ohmic contacts $T_{\rm c} \sim T_{\rm f} \gg T_{\rm in}$ one can replace the correlation function at temperature $T_{\rm in}$ in the leading order with zero temperature case and introducing the dimensionless variable of integration, $x=\pi t T_{\rm f}$, we get 
\begin{equation}
\label{Integral I1 in case of hot Ohmic contacts}
\mathcal{I}_1=T_{\rm f}\int \frac{(-1/2) dy}{(y-i0)\sinh(y-i0)}=T_{\rm f} \ln{2},
\end{equation}
and the sub-leading correction is proportional to $T_{\rm f} (T^2_{\rm in}/T^2_{\rm f}) \ll T_{\rm f}$. The second integral is given by 
\begin{equation}
\label{Integral I2 in case of hot Ohmic contacts}
\mathcal{I}_2=\frac{1}{T_{\rm f}C_{\rm f}} \lim_{\omega \to 0} [S^{\rm neq}_1(\omega)-S^{\rm eq}_{\rm f}(\omega)]\int \frac{x^2 dx}{2(x-i0)\sinh(x-i0)} = \frac{\pi^2}{8 T_{\rm in} R_{\rm q} C_{\rm f}}(T_{\rm c}+T_{\rm in}-2T_{\rm f}).
\end{equation}

%\begin{thebibliography}{99}
	
%\bibitem{}

%\end{thebibliography}